\documentclass[journal,comsoc]{IEEEtran}
\usepackage[T1]{fontenc}

\usepackage{amssymb,amsmath,amsfonts,amsthm}
\usepackage{mathrsfs}
\usepackage{cite}
\usepackage{array}
\usepackage{tabularx}
\usepackage[table]{xcolor}
\usepackage{color}
\usepackage{mathtools}
\usepackage{caption}
\usepackage{subcaption}
\usepackage{mathtools}
\usepackage{tikz,pgf}
\usetikzlibrary{calc,positioning,mindmap,trees,decorations.pathreplacing}
\usepackage{graphicx}
\usepackage{bbm}
\usepackage[utf8]{inputenc}
\usepackage{algorithm}

\makeatletter
\def\endthebibliography{%
	\def\@noitemerr{\@latex@warning{Empty `thebibliography' environment}}%
	\endlist
}
\makeatother

\begin{document}
	\title{Over-the-Air Computation for Distributed Systems: Something Old and Something New}
	
	\author{Zheng Chen, Erik G. Larsson, Carlo Fischione, Mikael Johansson, and Yura Malitsky
		\thanks{Zheng Chen and Erik G. Larsson are with the Department of Electrical Engineering, Link\"{o}ping University, Sweden (email:\{zheng.chen, erik.g.larsson\}@liu.se); \par  Carlo Fischione is with Digital Futures research center at Royal Institute of Technology (KTH), Sweden (email:carlofi@kth.se); \par Mikael Johansson is with the Department of Electrical Engineering and Computer Science, KTH, Sweden (email:mikaelj@kth.se); \par Yura Malitsky is with the Department of Mathematics, Link\"{o}ping University, Sweden (email:yurii.malitskyi@liu.se).}
	}
	
	\maketitle

	\begin{abstract}	
		Facing the upcoming era of Internet-of-Things and connected intelligence, efficient information processing, computation, and communication design becomes a key challenge in large-scale intelligent systems. Recently, Over-the-Air (OtA) computation has been proposed for data aggregation and distributed computation of functions over a large set of network nodes. Theoretical foundations for this concept exist for a long time, but it was mainly investigated within the context of wireless sensor networks. There are still many open questions when applying OtA computation in different types of distributed systems where modern wireless communication technology is applied. In this article, we provide a comprehensive overview of the OtA computation principle and its applications in distributed learning, control, and inference systems, for both server-coordinated and fully decentralized architectures. Particularly, we highlight the importance of the statistical heterogeneity of data and wireless channels, the temporal evolution of model updates, and the choice of performance metrics, for the communication design in OtA federated learning (FL) systems. Several key challenges in privacy, security, and robustness aspects of OtA FL are also identified for further investigation.
	\end{abstract}

	\section{Introduction}
	Wireless communications networks, designed with communication among humans in mind during the last fifty years, are now becoming
	overwhelmed with data traffic that stems from emerging artificial
	intelligence (AI) and machine learning (ML) applications. 
	One distinction of the data traffic required by AI/ML applications is that
	it is subject to entirely different performance requirements than
	traditional human-perceived data (e.g., text messages, voice calls or multimedia content).
	More explicitly, the traffic generated by AI/ML applications often contains computation results, which may be useful even if they are only received
	\emph{approximately correctly}.  
	Traditional quality-of-service performance metrics in communications such as error probability and rate may not be relevant for AI/ML applications. Rather, entirely different criteria will matter more, such as convergence speed of learning models, and statistical inference performance on average, in expectation over an ensemble of random data.  
	This shift will require a re-design of the networks, spanning from the physical layer to the application layer. It also will require going beyond classical Shannon information theory, which has underpinned all generations of networks so far.

	Within the AI/ML field, several classes of distributed algorithms have been developed with which autonomous agents (clients), that could be mobile users or sensors, collaboratively solve large-scale inference and learning tasks. The progress on collaborative and distributed intelligent systems has motivated the concept of wireless edge intelligence, which means that the computation tasks that were traditionally performed in centralized cloud servers are moved towards the wireless network edge. Implementing  distributed algorithms in wireless networks often requires iterative exchange and aggregation of information among the agents over resource-constrained communication links.
	One main challenge is the resource allocation among the agents, where the orthogonal division of frequency/time resources can be inefficient when the number of agents is very large.
	If we consider that the goal of communication for machine-perceived data differs from human-perceived data, joint computation and communication design without error-free bits has the potential to achieve better performance than classical digital transmission designs.
	
	Over-the-Air (OtA) computation, also referred to as AirComp, has emerged recently as a promising solution for computing functions of data from distributed nodes over wireless links by exploiting direct signal superposition in the analog domain \cite{ota-iot}. 
	In this article, we provide an overview of the basics, applications and research questions of OtA aggregation in distributed systems. 
	As compared to existing overviews in the literature, we summarize previously overlooked or under-explored aspects of OtA FL, such as statistical heterogeneity and temporal evolution of model updates, choices of performance metrics, and the communication design for fully decentralized systems.

	The organization of this article is summarized as follows. In Sec.~\ref{sec:tool}, we briefly explain the concept of OtA aggregation for distributed computation of nomographic functions. In Sec.~\ref{sec:application}, we present applications of OtA computation in distributed learning, control and inference systems. Then, we focus on the communication and signal processing aspects of OtA FL in Sec.~\ref{sec:comm-sp-ota}, and highlight the challenges in privacy and security aspects in Sec.~\ref{sec:privacy}. At last, in  Sec.~\ref{sec:conclusion}, we conclude the article and summarize important research directions.

	\section{Over-the-Air (OtA) Aggregation as a Computational Tool}
	\label{sec:tool}
	With OtA computation, by virtue of the superposition
	principle of the wave equation that governs the wireless
	multiple-access channel, information sent by the agents naturally adds
	up ``in the air''. Suppose $f\colon\mathbb{R}^N\to
	\mathbb{R}$ is a \emph{nomographic} function
	\cite{nomographic_function}, such that
	\begin{equation}
		f(x_1,\ldots,x_N)=\phi\left(\sum_{i=1}^N\psi(x_i)\right)
	\end{equation}
	for some scalar
	functions $\phi$ and $\psi$. Then every agent $i$ can perform
	$\psi(\cdot)$ prior to transmission, the medium (wave superposition)
	performs the actual aggregation ($\sum_i$), and the receiver performs
	$\phi(\cdot)$.  Any nomographic function can be computed this way
	through the OtA mechanism. Commonly used nomographic functions include arithmetic mean, geometric mean, and Eucliean norm.
	
	OtA aggregation makes a clean break with the notion that modern data
	transmission should be digital. Its principal benefit is that rather
	than giving the agents orthogonal resources (as in traditional digital
	communications), they send their data simultaneously, saving a factor
	of $N$ in spectrum resources.
	
	Though the core idea of OtA aggregation seems straightforward, in practice the picture is more complicated.  First,
	interference and noise (say $n$) will enter at the receiver, resulting in received post-processed signal as
	$\phi\left(\sum_{i=1}^N\psi(x_i)+n\right)$.  If the inversion carried out by
	$\phi$ is ill-conditioned, then the presence of $n$ may be
	detrimental. 
	Second, the basic OtA abstraction assumes that agents
	send their updates synchronously with perfect phase alignment. But in practice, some phase mismatches are unavoidable due to imperfectly known propagation delays and hardware impairments. The effect of imperfect synchronization needs to be accounted for in the reconstruction of aggregated model updates, and such aggregation errors are structurally different from other sources of ``noise'' in traditional ML systems. While in theory the synchronization issue can
	be resolved using calibration protocols, these are difficult to
	implement in emerging Internet-of-Things devices that must be low-cost
	and highly energy-efficient, and may operate at $>30$ GHz
	carrier frequencies where phase drift is a major issue.
	
	\section{Applications of OtA Computation in Distributed Systems}
	\label{sec:application}
	Theoretically, OtA computation can be useful in any distributed system that requires combining and aggregating data from multiple sources at one common receiver without the need of receiving each independent stream correctly. In this section, we provide some application examples and highlight the relation between the accuracy of data aggregation and the underlying tasks/applications performed by the system.

	\begin{figure*}[t!]
		\vspace{-0.2cm}
		\centering
		\begin{subfigure}[b]{0.35\textwidth}
			\centering
			\includegraphics[width=\textwidth]{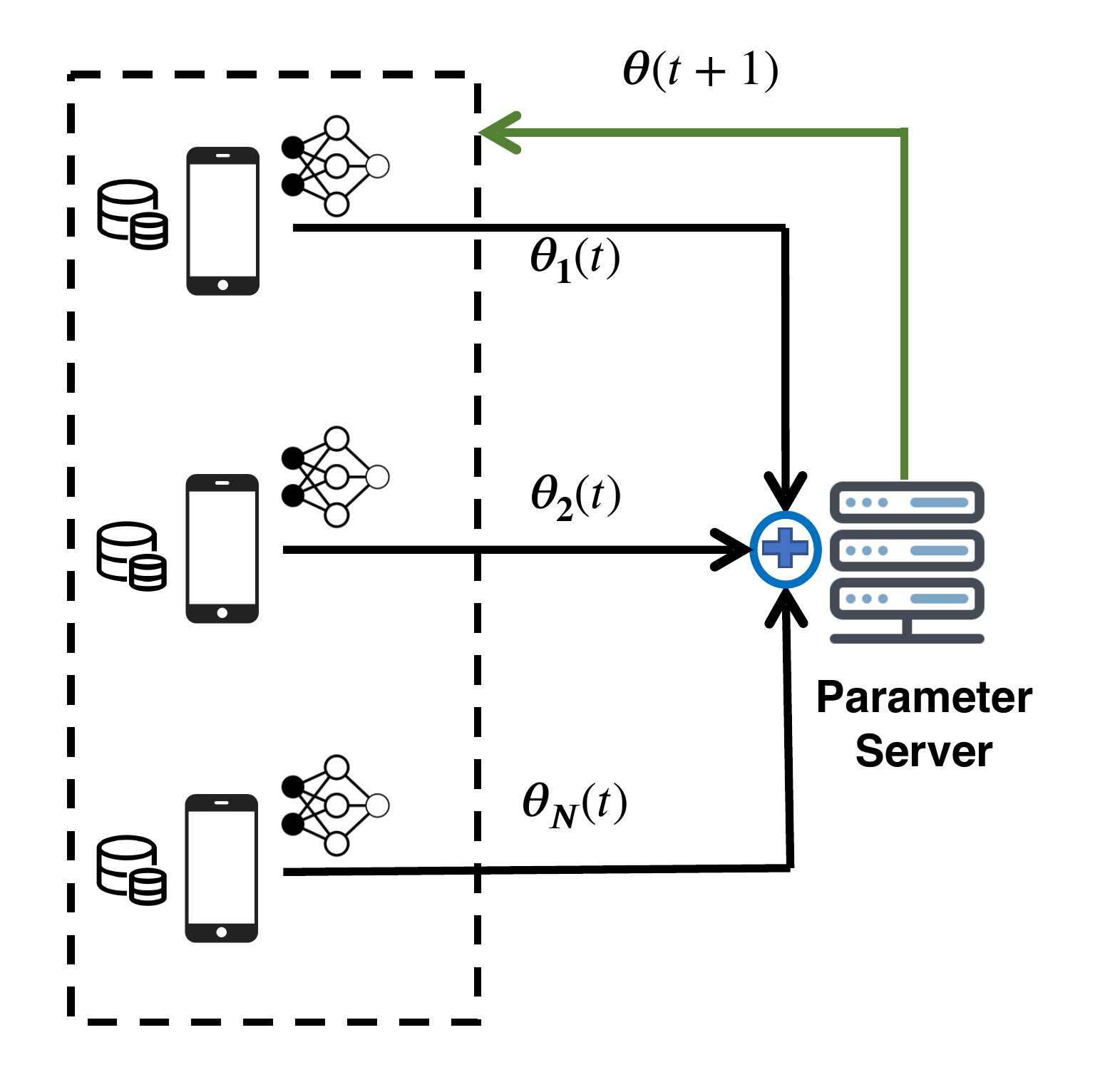}
			\caption {Server-coordinated}
		\end{subfigure}
		\begin{subfigure}[b]{0.35\textwidth}
			\centering
			\includegraphics[width=\textwidth]{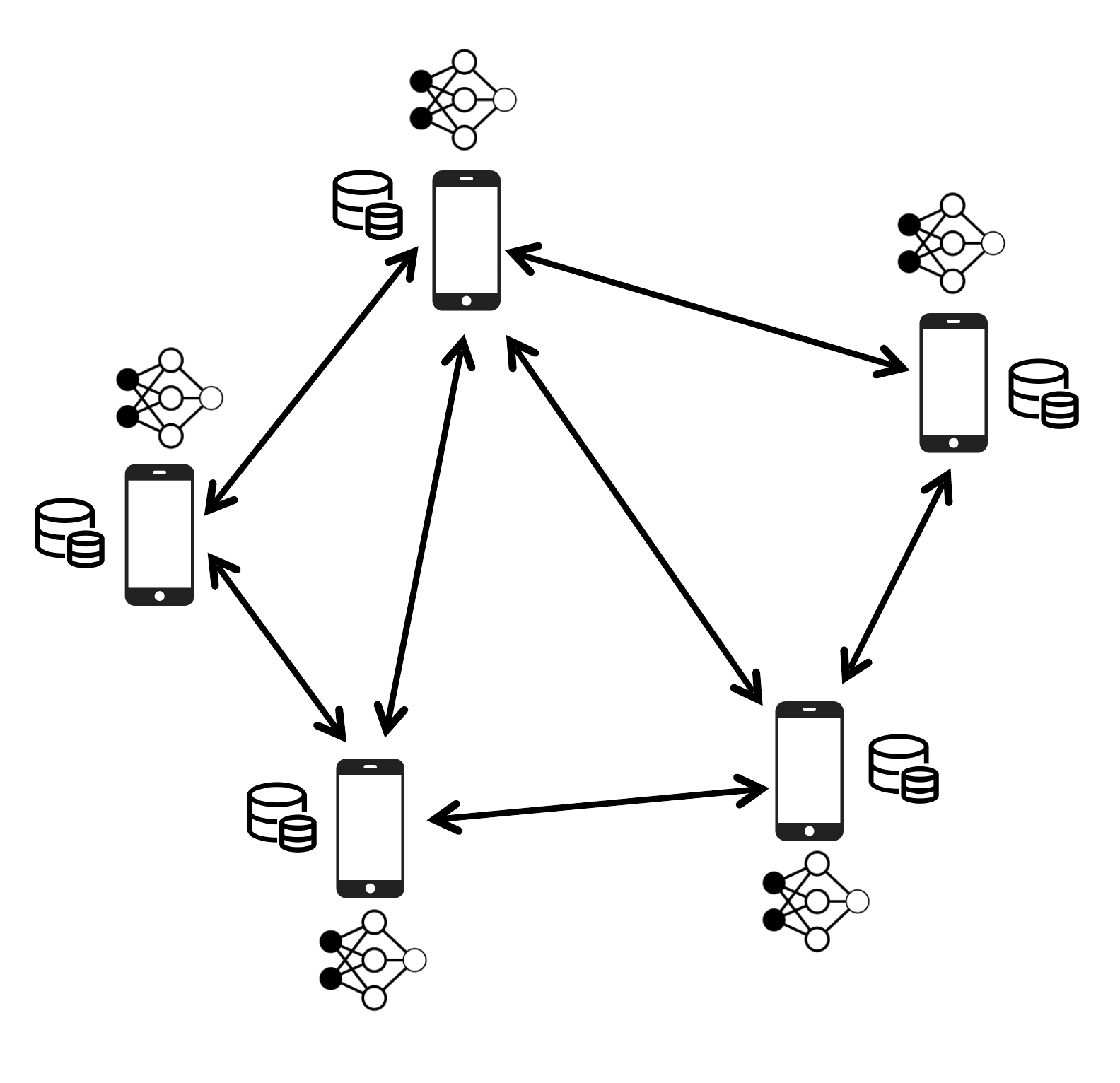}
			\caption{Fully decentralized}
		\end{subfigure}
		\caption{System architectures of server-coordinated and fully decentralized collaborative ML systems.}
		\label{fig:DML}
	\end{figure*}	
	
	\subsection{Distributed ML with Collaborative Training}
	Currently, most research activities on the application of OtA computation focus on distributed ML with multiple agents collaborating in training a common ML model, as shown in Fig.~\ref{fig:DML}. The most outstanding example is server-coordinated FL with a master-worker architecture, where the computation objective at the server is to obtain the weighted sum of the model update vectors from the agents after each round of local training. Collaborative ML can also be implemented in fully decentralized systems, where information exchange and fusion take place between locally connected agents without communication with a central server. This part will be elaborated with more details in Sec.~\ref{sec:fully decentralized}.\footnote{In some existing literature, these two architectures are referred to as ``centralized FL'' and ``decentralized FL'', respectively.}
	
	\begin{figure}[h!]
		\centering
		\includegraphics[width=\linewidth]{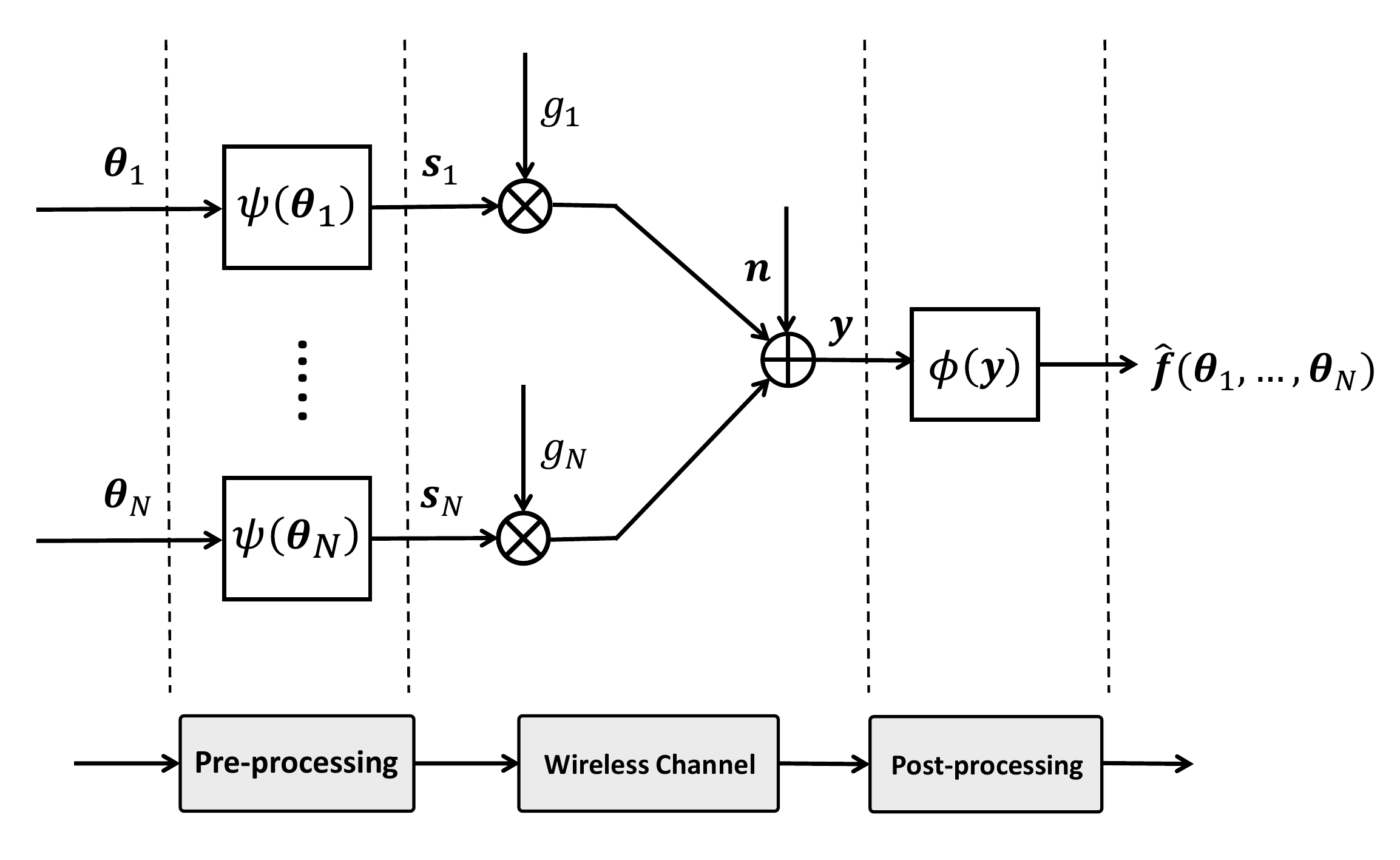}
		\caption{Illustration of OtA computation over a multiple access fading channel. $\boldsymbol{\theta}_i$ is the model update vector from user $i$, $g_i$ is the channel gain between the $i$-th user and the server, $n$ is the additive noise. $\psi$ and $\phi$ refer to the pre-processing function at the sender (user) side and the post-processing function at the receiver (server) side, respectively.}
		\label{fig:ota}
	\end{figure}

	In server-coordinated FL systems, to effectively compute the weighted average of the model updates at the server, each agent pre-processes its data 
	to compensate for the amplitude degradation and phase rotation in the wireless channel, subject to some power constraints. At the server side, the received superimposed signal is post-processed to obtain an estimate of the function to be computed. This procedure is illustrated in Fig.~\ref{fig:ota}.

	\subsection{Distributed Control Systems}
	Distributed control systems rely on multiple geographically distributed sensors and actuators to control a larger system, such as a factory or an infrastructure network. Increasingly often, the control loops are closed over wireless communication links. OtA computation could potentially improve the efficiency of such systems. For example, \cite{ota-control} considers a simple scenario where a single control loop with a single actuator uses measurements from multiple sensors.
	OtA aggregation is applied such that all sensors simultaneously send scaled signals containing the measured plant state to the actuator. The actuator receives the superimposed control signals in the presence of channel noise. The scaling factors at the sensor and the actuator sides can be designed in a way to minimize the distortion of the received control signal, which eliminates the need of a control unit. Thus the sensor-to-actuator feedback occurs without a separate unit.  The study in \cite{ota-control} is arguably just the initial application of OtA to distributed control, and thus several questions remain to be answered in this setting, such as the relation between system stability and the distortion of aggregated signal, and the impact of output feedback and transmit power limitations on the system performance.

	\subsection{Distributed Estimation and Inference}
	One example in this application is to estimate some random field with distributed sensors when the dimension of the observation at each sensor is much smaller than the dimension of the field \cite{distributed-inference}. 
	This can be done by applying distributed linear estimation with consensus $+$ innovations algorithms for the information exchange and fusion among cooperative sensors following an iterative process. In every iteration, the exchanged information contains the current estimate of the random field, where OtA can be applied for data aggregation. The impact of communication constraints and costs on the reconstruction of the random field is an interesting direction to explore.

	There are numerous potential OtA applications that fall within the general framework of federated analytics. The most important characteristics of these applications are: 1) the communication goal is to compute \textit{functions of distributed data}; 2) the computation result only needs to be \textit{approximately correct}. In other words, not all distributed systems can potentially benefit from OtA computation for data aggregation. Currently, most articles on the application of OtA computation focus on server-coordinated FL. Therefore, in the following sections, we focus on communication design and signal processing in OtA FL. Rather than summarizing existing work on this topic, the purpose of this article is to point out what is missing in the current literature and which aspects of OtA FL should be further investigated. 
	
	\section{Communication and Signal Processing Perspective of OtA FL}
	\label{sec:comm-sp-ota}
\begin{figure}
		\vspace{-0.2cm}
		\centering
		\begin{subfigure}[b]{0.5\textwidth}
			\centering
			\includegraphics[width=\textwidth]{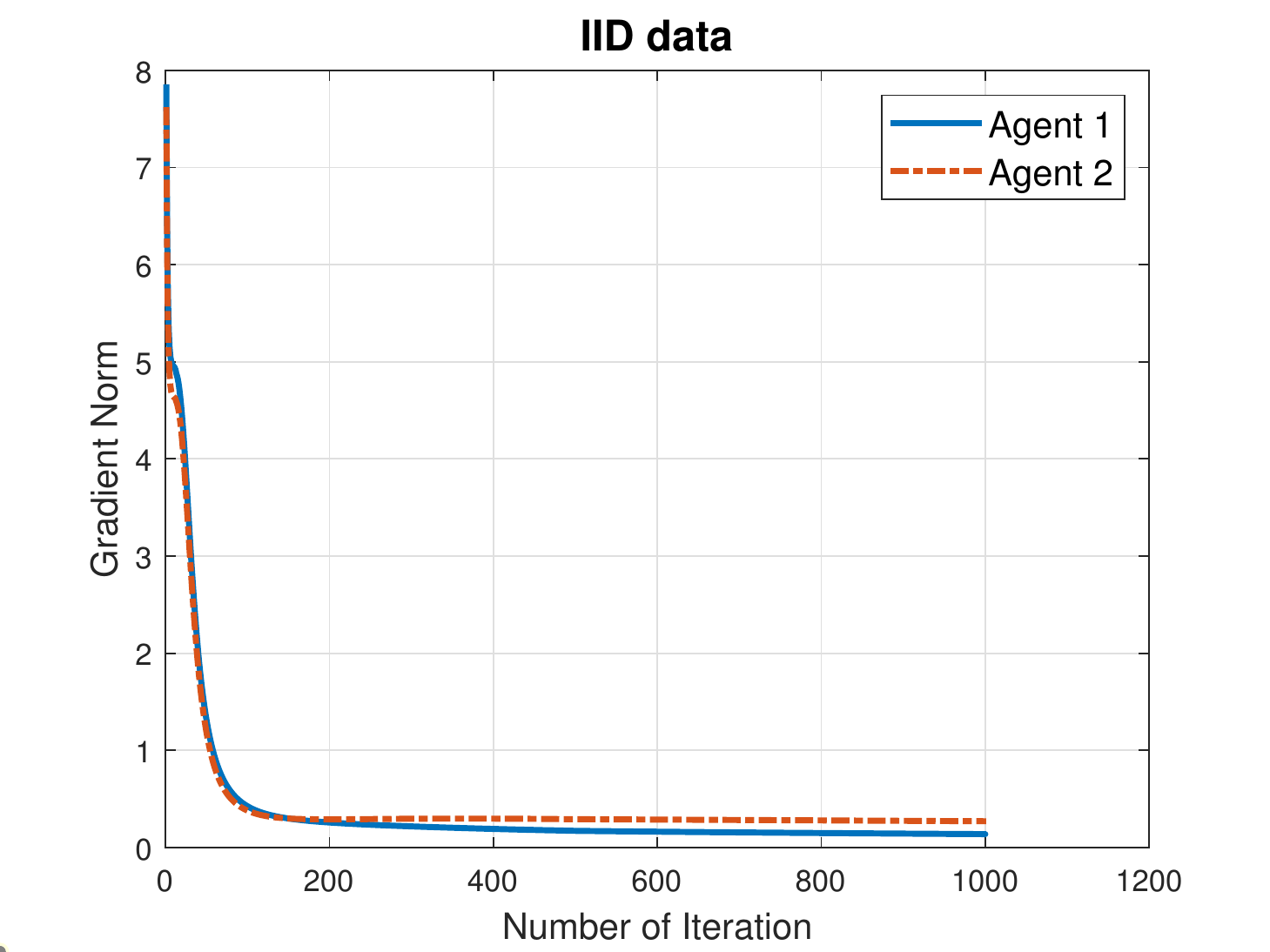}
			\caption {IID data}
		\end{subfigure}
		\begin{subfigure}[b]{0.5\textwidth}
			\centering
			\includegraphics[width=\textwidth]{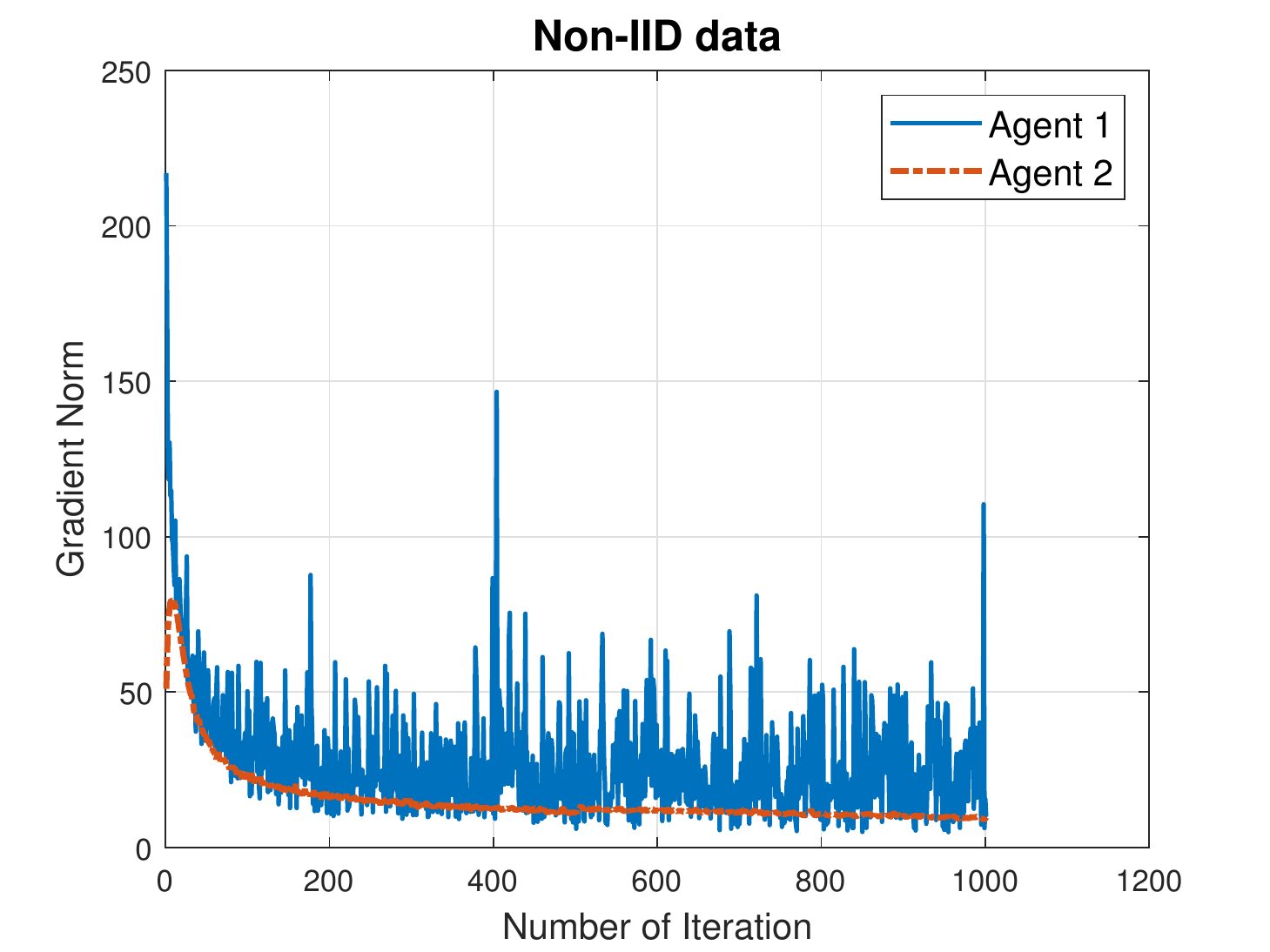}
			\caption{Non-IID data}
		\end{subfigure}
		\caption{Evolution of the gradient norm by training a convolutional neural network for a classification problem. MNIST dataset is used and the data samples are evenly distributed over $100$ agents divided into two groups. With IID data, all agents have training samples containing all digits. With non-IID data, one group (which Agent $1$ belongs to) has only samples with digits $0-4$, and the other group (which Agent $2$ belongs to) has only $5-9$.} 
		\label{fig:gradient-norm}
	\end{figure}

	Even though the main advantage of OtA computation for wireless data aggregation is the efficient usage of communication resources, there are still many factors that can affect the computation and communication efficiency in OtA systems. In the existing literature on OtA FL, the following aspects have been either ignored or under-explored:
	\begin{itemize}
		\item the impact of non-IID data on the statistical heterogeneity of the local model updates;
		\item the disparity between the channel conditions of different agents and its impact on the statistics of the reconstruction error;
		\item the temporal evolution/correlation of the model updates (or gradient vectors) along the communication rounds;
		\item other performance metrics than the mean-square error (MSE) of the aggregated model updates.
	\end{itemize}
	One example is that many existing works assume that the model updates from different agents follow the same distribution with unit variance. This inherently suggests that all the agents in the system are statistically identical. As illustrated in Fig.~\ref{fig:gradient-norm}, this assumption is generally not true with non-IID training data across different agents.\footnote{Note that the obtained gradient norms depend on many parameters in the simulation setting, such as data distribution, learning rate, and batch size. In Fig.~\ref{fig:gradient-norm} with non-IID data, Agent 2 has less fluctuation in the gradient norm evolution than Agent 1, which implies that the optimal model parameters obtained with local data at Agent 2 might be closer to the global optimal model. However, this simulation result does not necessarily mean that Agent 1 always have worse training performance than Agent 2.} Another observation from Fig.~\ref{fig:gradient-norm} is that the gradient norms evolve over time. Only a few existing papers have considered the temporal structure of model updates and its impact on the data compression design \cite{ozfatura2021time, fan2021temporal}. The joint consideration of these aspects mentioned above brings new challenges and questions in OtA FL that differ from a classical statistical estimation setting.

	\subsection{Computation, Communication and Convergence}
	\label{sec:comm_conv}
	An important aspect of FL over wireless links is the interplay between computation and communication costs and the impact of radio resource management on the convergence performance. The design space of resource allocation in wireless FL is vast, but the fundamental question is: under given constraints on frequency/time/power, how can we allocate these resources to different agents in every communication round such that the FL algorithm converges as fast and as accurately as possible? 
	Computational complexity is another factor that can affect the convergence performance and energy consumption. 
	Consequently, resource limitations pose constraints on the number of agents that can participate in each communication round and the power allocation between computation (local training) and communication (sending model updates). 
	
	In the digital domain, work on communication-efficient ML has centered
	around compression techniques such as quantization and sparsification of gradient vectors. For FL over rate-limited wireless networks, scheduling and resource (frequency, time, power) allocation can be optimized to improve the system convergence performance under the limitation of communication resources.
	In the OtA domain, since all users share the same frequency/time resources, the communication efficiency is mostly affected by the data compression and scheduling design. Particularly, sparsification plays an important role in reducing the dimension of the transmitted data, thus reducing the consumed radio resources. For this to work, the locations of the sparsified elements need to the sent as side information to the receiver through some reliable channels.
	Furthermore, we can explore linear compression techniques inspired by compressed sensing method to recover the aggregated model parameter vector at the server side.

	In the digital domain, the impact of scheduling lies in the fact that with more devices participating in model uploading, less frequency/time resources are allocated to each user, which results in higher compression loss that may deteriorate the accuracy of the aggregated model. With OtA aggregation, scheduling is also important because all the agents apply a common scaling factor in the pre-processing step such that the transmitted signals satisfy the power constraints at all agents. This scaling factor affects the variance of the effective noise after post-processing at the receiver side. From the perspective of minimizing the reconstruction error, it might be beneficial to drop several bottleneck users with large gradient norms and bad channel conditions. From the perspective of improving convergence by reducing the variance of the aggregated model, it is crucial to include as many users as possible in the model aggregation. The interplay between computation, communication and convergence is a critical design question of OtA FL.

	\subsection{Temporal Statistics and Adaptive Learning}
	For distributed ML over wireless networks, the temporal evolution of the generated model updates (gradient vectors) and the communication link conditions is another fresh aspect that makes the OtA problem more interesting. 
	
	For gradient vectors in FL, there is some evidence of a predictable structure \cite{fan2021temporal}, in that the amplitude of the different gradient elements changes slowly over time. This inherently suggests that the impact of communication resources on the system performance is not equal over time. Additionally, some recent work on FL has shown that the convergence in later iterations is more sensitive to large compression loss than the earlier ones. This observation has motivated the consideration of adapting communication resources along the iterations such that the aggregated model is received with higher accuracy during the later iterations. 
	This fact was exploited in~\cite{khirirat2021flexible} to derive an adaptive gradient compression mechanism that maximizes the optimization progress (reduction in expected loss function value) per communication resource (\emph{e.g.}, channel or power) used in every iteration. Similar ideas can be explored for OtA FL, where energy allocation and scheduling decisions can be adapted along the iterations to achieve faster convergence under a total energy budget for the entire training process. In this case, the channel statistics and data statistics must be jointly accounted for when making transmission decisions. The development of adaptive scheduling and resource allocation policies for iterative learning algorithms with OtA computation is a challenging open problem.

	\subsection{OtA Computation with Multiple-Antenna Technology}
	Most published research on OtA FL considers the case with single-antenna agents and servers. Given the enormous success of multiple-antenna technology in cellular communications, especially  massive MIMO, it is conceivable that the use of multiple antennas would greatly benefit also OtA FL setups.
	A crucial aspect of OtA FL is that the agents need to know, on a per-antenna basis, the physical channel response in order to pre-compensate for the channel phase-rotation and amplitude attenuation, before transmission. 
	This is required such that the signals from all agents can be added up constructively, in-phase at the server.  But with multiple antennas at the server, there will be no universal phase-rotation per agent antenna that works for all server antennas. 
	
	Two basic cases can be distinguished. Either the agents can obtain knowledge of the per-antenna channel responses (including their phase), or they cannot.
	If these channel responses are available, then the agents can optimize their transmission, which includes selecting beam patterns (in case of multiple agent antennas) and phase-rotations, such that the weighted sum of the transmitted  gradient updates can be inferred from the signals collectively received at the server antennas.
	In contrast, if the channel responses are unknown, then the agents cannot meaningfully pre-compensate, or even beamform, their transmissions to the server and one has to either resort to ``blind'' OtA aggregation techniques \cite{amiri2021blind}, or to spatial multiplexing of the gradients using a conventional multiple-antenna decoding algorithm such as zero-forcing.
	The latter option allows for individual sparsification and compression of the gradients, and it was shown in \cite{becirovic2022optimal} that this approach is superior as long as the number of antennas across all agents is smaller than at the server and the channel coherence allows for the transmission of individual uplink reference signals (which are required to separate the spatially multiplexed streams).

	The design of optimal transmission schemes for multi-antenna agents and servers remains an open research problem, whose solution entails an accurate modeling of the effects of channel coherence, channel estimation errors, gradient compression, and calibration -- which eventually determines to what extent the channel response phases can be obtained by the agents.
	It also requires the use of appropriate performance metrics, that go beyond the metrics conventionally used in the design of multiple-antenna communication systems.

	\subsection{Other Metrics than the MSE of Aggregated Data}
	Many existing research on OtA FL uses the MSE of the aggregated model updates as the performance metric to optimize the communication (scheduling and power control) design. From this perspective, the formulation of the optimization problem becomes independent of the ML setting. There is no clear evidence that minimizing the MSE (without optimal re-scaling) is equivalent to optimizing the learning performance. From the learning perspective, the direction of the aggregated gradient vector matters more than the magnitude. The impact of OtA aggregation error on the system performance cannot be simply measured by the squared norm of the difference between transmitted and received signal vectors.
	
	Additionally, it is shown in \cite{wei2022federated} that the channel noise does not have a significant impact on the convergence performance if the channel noise does not dominate the noise in stochastic gradient vectors caused by random data sampling. Based on this remark, we expect that the effect of OtA estimation error on the learning performance will depend on the level of noise in the computation of stochastic gradients.  
	Another concern with using the MSE as the optimization objective is the impact of scheduling. As briefly explained in Sec.~\ref{sec:comm_conv}, minimizing the MSE of aggregated data will motivate us to drop many bottleneck agents, which in turn increases the bias of the aggregated model, especially under the non-IID training data setting. In Table~\ref{tab:MSE-accuracy}, we show an example of the accuracy-MSE tradeoff in OtA FL. This result suggests that the ``plain'' MSE (without proper re-scaling) of the aggregated signals from the set of scheduled users does not fully reflect how close the aggregated gradient vector is to the true gradient. An efficient scheduling design needs to balance the bias caused by scheduling users with heterogeneous data and the estimation error of OtA aggregation under the power constraints.
	\begin{table*}[t]
		\centering
		\renewcommand{\arraystretch}{1.2}
		\begin{tabular}{|c|c|c|c|c|c|c|}
			\hline   
			\text{Scheduled users}&    $50$&  $40$&  $30$& $20$ & $10$\\
			\hline
			\text{Test accuracy}&    $92\%$&   $71.65\%$& $56.86\%$ & $42.45\%$ & $27.37\%$\\
			\hline
			\text{MSE}&    $7.75e-14$  &$3.70e-15$& $7.54e-17$ &$7.60e-18$ &$1.81e-18$\\
			\lasthline
		\end{tabular}
		\caption{Test accuracy (after $100$ communication rounds) vs. MSE of aggregated model updates for different numbers of scheduled users. The total number of users is $50$. The entire training dataset (MNIST) is equally divided among the users but each user only contains data samples with at most $2$ digits (non-IID data setting). The scheduling decision is made by minimizing the MSE of the aggregated model updates from the set of scheduled users. }
		\label{tab:MSE-accuracy}
	\end{table*}

	The same consideration can be extended to many other applications of OtA in distributed systems, such as the ones mentioned in Section \ref{sec:application}. Finding meaningful performance metrics for OtA computation in different distributed systems remains an important open question. 
	
	\subsection{OtA Computation for Fully Decentralized Systems}
	\label{sec:fully decentralized}
	
	With fully decentralized ML, the system only relies on local computations and information exchange between locally connected nodes to solve a decentralized optimization problem. Although several conceptual algorithms have been proposed for such systems, such as distributed sub-gradient method and distributed CoCoA, using OtA to perform such computations has rarely been investigated. The convergence speed of such algorithms is affected by the connectivity of the network nodes over time \cite{Rabbat}, which in turn is affected by how the scheduling of OtA nodes is performed in every communication round. This raises new challenges in terms of joint computation-communication methods, stability, and optimality of the algorithms.

	The first challenge is the communication scheduling problem. Here, as opposed to the usual “many-to-one” topology, fully decentralized learning requires “many-to-many” communication over unreliable wireless links with limited range, where each node needs to send and aggregate information at the same time. Thus, the communication protocol for information broadcasting, reception and fusion will play a major role in the performance (optimality and convergence) and scalability of ML algorithms. Should the protocol be synchronous where model updating occurs after all agents receive information from their neighbors, or asynchronous as in gossip algorithms? 
	There are many open questions when considering the broadcast nature and superposition of signals in OtA channels. 
	Randomized communication protocols that combine features of distributed medium access, spatial reuse of communication resources and compressed communication could be particularly attractive to develop.

	Another challenge is the “precoding problem”. Normally, an agent adapts the phase and amplitude of its message before transmission so that the OtA channel returns the desired computation at a specific receiver. But in a fully decentralized system, the OtA channel must simultaneously return many desired computation results at different nodes despite the fact that we can control only one precoder at each user. How to select one precoder that can return several OtA computations at the same time? One possible solution is to combine the degrees of freedom in space (joint precoding and decoding at different nodes) and in time (coding over multiple slots) \cite{space-time-ota}. Other designs can be developed by following the methods of distributed estimation theory. In such an estimation set-up, the nodes of a wireless network cooperate to perform a distributed estimation process, where each node broadcasts its local estimate to the neighbors, which in their turn, use such received estimate to update their local estimates. The main challenge in this case is that the channel quality and dynamics in OtA channels can threaten the stability and convergence of the ML algorithms if they would merely follow the distributed estimation algorithms.

	\section{Privacy, Security and Robustness of OtA FL}
	\label{sec:privacy}
	In general, distributed learning systems can be vulnerable to different types of attacks, depending on which level of information is shared among the agents in the system \cite{ma2022trusted}. In this article, we focus on OtA FL with transmissions of shared training model parameters over wireless links. The specific features of aggregating model updates through OtA computation generate new perspectives of potential threats and possible defense methods that are unique in OtA systems.

	\subsection{Privacy and Security}	
	One of the main advantages of FL is to protect data privacy by only sharing model parameter updates instead of the raw training data with the central server. However, FL can still be vulnerable to privacy and security attacks, such as membership inference attack and model poisoning attack. As compared to FL with digital transmission of model updates, one distinct feature of OtA FL is that channel noise can be exploited as a natural random perturbation function for improving differential privacy \cite{privacy-free}. Also, the server only receives the aggregated signals from the agents without being able to decode each of the data stream, which makes it difficult to infer information about the local data samples at each agent. 
	
	Nonetheless, existing methods for improving differential privacy are mostly built on the idea of gradient obfuscation (e.g., adding random perturbations to the gradient vectors), which leads to an accuracy-privacy tradeoff since the added perturbations in the aggregated gradient can compromise the learning performance. One possibility is to apply \emph{spatially correlated} perturbations across different agents such that the aggregated perturbations add up to zero at the server side \cite{privacy-ota-fl}. In this direction, many other ideas can be developed, such as adaptive privacy budget allocation and long-term optimization of privacy-accuracy tradeoffs.

	\subsection{Robustness and Reliability}
	Distributed SGD algorithms are generally vulnerable to Byzantine attacks, where the goal of attack is to prevent the system from converging. Conventional gradient aggregation techniques that are Byzantine fault-tolerant, typically rely on distance-based clustering and
	truncation schemes to eliminate the influence of outliers that may be
	Byzantine attackers. 
	With OtA aggregation, in contrast to with digital transmission
	schemes, the different data streams are inseparable. Because of the
	inherent anonymity of OtA, it is in principle easy for an adversary to
	send malicious messages (e.g, false gradients) and harm the computations. Any random perturbation signal can affect directly the aggregated model update, and the server will not have the ability to distinguish whether the received model update has been ``poisoned''. Such attacks are known as model poisoning. However, the defense strategies
	proposed in the literature such as \textit{Krum}, \textit{geometric median}, and \textit{Bulyan} require exact knowledge about each individual
	gradient, which is not directly possible for OtA. 
	
	In addition to model poisoning, it is well-known that carefully designed adversarial examples can degrade the trained model. These are inputs that are virtually indistinguishable from the original input to a human, but on which the trained model shows drastically different results. To circumvent that, the prevailing approach is adversarial training, which is often formulated as a minimax optimization problem, where the adversarial examples and the model weights are alternatively updated. Such learning typically requires many more epochs, and hence, more communication rounds. Because of the latter, the advantages of applying OtA computation instead of digital communication design are obvious. However, the robustness of adversarial training with OtA computation remains to be investigated.
	
	In general, very little research has been performed on robust model update aggregation in OtA FL. How to achieve robust and reliable computation will be a main challenge and open question for OtA-assisted distributed systems.

	\section{Conclusions}
	\label{sec:conclusion}
	As we enter the 6G era, future wireless systems need to support efficient information processing and data aggregation in large-scale distributed intelligent systems. OtA computation shows a clear advantage in resource efficiency as compared to traditional digital communication with separated source and channel coding schemes. Though the concept of distributed computation of functions over multiple access channels is not new, there are still many new challenges and open questions that are worth investigating. In this article, we provided an overview of the theory and applications of OtA computation, and highlighted the importance of the following aspects:
	\begin{itemize}
		\item The statistical heterogeneity of users (both in data distribution and channel condition) and temporal evolution of model updates will have a strong impact on communication efficiency. This is often overlooked in the existing literature on OtA FL.
		\item Finding appropriate performance metrics for different types of distributed systems with OtA computation is crucial for optimizing the resource efficiency.
		\item Many open questions remain in the communication and computation design for OtA in fully decentralized systems with local information exchange.
		\item OtA FL is vulnerable to model poisoning attacks since any perturbation signal can directly affect the aggregated model update. The robustness and reliability of OtA systems will be a great challenge for its application in real-world networks.  
	\end{itemize}

	\section*{Acknowledgment}
	The authors would like to give acknowledgment to Chung-Hsuan Hu, doctoral student at Link\"{o}ping University (Sweden), and Jeremy Weill, Master student at EPFL (Switzerland), for their contributions with the simulation results presented in the article.


\begin{thebibliography}{10}
		\providecommand{\url}[1]{#1}
		\csname url@samestyle\endcsname
		\providecommand{\newblock}{\relax}
		\providecommand{\bibinfo}[2]{#2}
		\providecommand{\BIBentrySTDinterwordspacing}{\spaceskip=0pt\relax}
		\providecommand{\BIBentryALTinterwordstretchfactor}{4}
		\providecommand{\BIBentryALTinterwordspacing}{\spaceskip=\fontdimen2\font plus
			\BIBentryALTinterwordstretchfactor\fontdimen3\font minus
			\fontdimen4\font\relax}
		\providecommand{\BIBforeignlanguage}[2]{{%
				\expandafter\ifx\csname l@#1\endcsname\relax
				\typeout{** WARNING: IEEEtran.bst: No hyphenation pattern has been}%
				\typeout{** loaded for the language `#1'. Using the pattern for}%
				\typeout{** the default language instead.}%
				\else
				\language=\csname l@#1\endcsname
				\fi
				#2}}
		\providecommand{\BIBdecl}{\relax}
		\BIBdecl
		
		\bibitem{ota-iot}
		G.~Zhu, J.~Xu, K.~Huang, and S.~Cui, ``Over-the-air computing for wireless data
		aggregation in massive {IoT},'' \emph{IEEE Wireless Communications}, vol.~28,
		no.~4, pp. 57--65, 2021.
		
		\bibitem{nomographic_function}
		M.~Goldenbaum, H.~Boche, and S.~Stańczak, ``Nomographic functions: Efficient
		computation in clustered gaussian sensor networks,'' \emph{IEEE Transactions
			on Wireless Communications}, vol.~14, no.~4, pp. 2093--2105, 2015.
		
		\bibitem{ota-control}
		P.~Park, P.~D. Marco, and C.~Fischione, ``Optimized over-the-air computation
		for wireless control systems,'' \emph{IEEE Communications Letters}, vol.~26,
		no.~2, pp. 424--428, 2022.
		
		\bibitem{distributed-inference}
		S.~Kar and J.~M. Moura, ``Consensus + innovations distributed inference over
		networks: cooperation and sensing in networked systems,'' \emph{IEEE Signal
			Processing Magazine}, vol.~30, no.~3, pp. 99--109, 2013.
		
		\bibitem{ozfatura2021time}
		E.~Ozfatura, K.~Ozfatura, and D.~G{\"u}nd{\"u}z, ``Time-correlated
		sparsification for communication-efficient federated learning,'' in
		\emph{2021 IEEE International Symposium on Information Theory (ISIT)}.\hskip
		1em plus 0.5em minus 0.4em\relax IEEE, 2021, pp. 461--466.
		
		\bibitem{fan2021temporal}
		D.~Fan, X.~Yuan, and Y.-J.~A. Zhang, ``Temporal-structure-assisted gradient
		aggregation for over-the-air federated edge learning,'' \emph{IEEE Journal on
			Selected Areas in Communications}, vol.~39, no.~12, pp. 3757--3771, 2021.
		
		\bibitem{khirirat2021flexible}
		S.~Khirirat, S.~Magn{\'u}sson, A.~Aytekin, and M.~Johansson, ``A flexible
		framework for communication-efficient machine learning,'' in
		\emph{Proceedings of the AAAI Conference on Artificial Intelligence},
		vol.~35, no.~9, 2021, pp. 8101--8109.
		
		\bibitem{amiri2021blind}
		M.~M. Amiri, T.~M. Duman, D.~G{\"u}nd{\"u}z, S.~R. Kulkarni, and H.~V. Poor,
		``Blind federated edge learning,'' \emph{IEEE Transactions on Wireless
			Communications}, vol.~20, no.~8, pp. 5129--5143, 2021.
		
		\bibitem{becirovic2022optimal}
		E.~Becirovic, Z.~Chen, and E.~G. Larsson, ``Optimal {MIMO} combining for blind
		federated edge learning with gradient sparsification,'' in \emph{2022 IEEE
			23rd International Workshop on Signal Processing Advances in Wireless
			Communication (SPAWC)}, 2022, pp. 1--5.
		
		\bibitem{wei2022federated}
		X.~Wei and C.~Shen, ``Federated learning over noisy channels: Convergence
		analysis and design examples,'' \emph{IEEE Transactions on Cognitive
			Communications and Networking}, 2022.
		
		\bibitem{Rabbat}
		A.~Nedić, A.~Olshevsky, and M.~G. Rabbat, ``Network topology and
		communication-computation tradeoffs in decentralized optimization,''
		\emph{Proceedings of the IEEE}, vol. 106, no.~5, pp. 953--976, 2018.
		
		\bibitem{space-time-ota}
		Z.~Chen and Y.~Malitsky, ``Over-the-air computation with multiple receivers: A
		space-time approach,'' \emph{arXiv preprint arXiv:2208.11751}, 2022.
		
		\bibitem{ma2022trusted}
		C.~Ma, J.~L. Wei, B.~Liu, M.~Ding, L.~Yuan, Z.~Han, H.~V. Poor \emph{et~al.},
		``Trusted {AI} in multi-agent systems: An overview of privacy and security
		for distributed learning,'' \emph{arXiv preprint arXiv:2202.09027}, 2022.
		
		\bibitem{privacy-free}
		D.~Liu and O.~Simeone, ``Privacy for free: Wireless federated learning via
		uncoded transmission with adaptive power control,'' \emph{IEEE Journal on
			Selected Areas in Communications}, vol.~39, no.~1, pp. 170--185, 2021.
		
		\bibitem{privacy-ota-fl}
		J.~Liao, Z.~Chen, and E.~G. Larsson, ``Over-the-air federated learning with
		privacy protection via correlated additive perturbations,'' in \emph{58th
			Annual Allerton Conference on Communication, Control, and Computing
			(Allerton)}, 2022, pp. 1--8.
		
	\end{thebibliography}

\end{document}